\documentclass[showpacs,preprintnumbers,amsmath,amssymb]{revtex4}
\usepackage{tabularx,graphicx}
\begin{document}




\newcommand{\beq}{\begin{equation}}
\newcommand{\eeq}{\end{equation}}
\newcommand{\beqn}{\begin{eqnarray}}
\newcommand{\eeqn}{\end{eqnarray}}
\newcommand{\bmath}{\begin{subequations}}
\newcommand{\emath}{\end{subequations}}
\title{Superconductors as giant atoms: qualitative aspects}

\author{J.E. Hirsch}
  \address{Department of Physics, University of California, San Diego\\ 
La Jolla, CA 92093-0319}

\begin{abstract} When the Fermi level is near the top of a band the carriers (holes) are maximally dressed by electron-ion
and electron-electron interactions.  The theory of hole superconductivity predicts that only in that case can
superconductivity occur, and that it  is driven by $undressing$   of the carriers at the Fermi energy upon pairing. 
Indeed,
experiments show that  dressed hole carriers in the normal state become
undressed electron carriers in the superconducting state.  This leads to a description of superconductors as giant atoms,
where undressed time-reversed electrons are paired and
 propagate freely in a uniform positive background. The pairing gap
provides rigidity to the  wavefunction, and electrons in the giant atom respond to magnetic fields the same way as
electrons in diamagnetic atoms. We predict that there is an electric field in the interior of superconductors and 
that the charge distribution is inhomogeneous, with higher concentration of negative charge near the surface;
that the ground state of superconductors has broken parity and possesses macroscopic spin currents,   and that negative charge spills out when a body becomes superconducting.
\end{abstract}

\maketitle


\section{Introduction}

 The Bloch-Landau description of metals is appropriate to describe the normal state. It describes
dressed quasiparticles (electrons or holes) whose charge sign depends on the location of the
Fermi level in the band, and whose effective mass is different from the free electron mass due to
electron-ion and electron-electron interactions. Instead, we propose that in
superconductors the carriers of the supercurrent are $bare$ $electrons$ and that superconductors
should be understood as giant atoms.

Unlike the conventional BCS-Eliashberg theory, the theory of hole superconductivity discussed here 
 is proposed to describe $all$ superconductors\cite{holesc}. It proposes that superconductivity arises only when
the Fermi level is near the top of an electronic energy band, and is driven by $undressing$\cite{undressing}
 of the carriers at
the Fermi energy upon pairing. It requires the existence of electron-ion and electron-electron interaction 
but $not$ of electron-phonon interaction, so that it predicts that superconductivity can occur in a solid even
if the ionic mass is infinite. On the other hand it predicts that superconductivity cannot occur if there are
no $antibonding$ $electrons$ at the Fermi energy, i.e. if the Fermi level is close to the 
bottom of the band\cite{holeelec}.

When carriers 'undress' they become more free-electron-like. We argue that certain experiments in
superconductors show that the superfluid carriers are in fact completely free-electron like. Hence they
have completely 'undressed' from both electron-ion and electron-electron interactions. 
We argue that carriers at the Fermi energy are most highly dressed by both the electron-electron and the
electron-ion interaction when the Fermi level is close to the top of the band. Furthermore, the 
formalism shows that the dressing
is highest when the ions are negatively charged\cite{hole93}. Naturally, the most favorable
situation for superconductivity is when the carriers are most heavily dressed in the normal state, because
that is when the most  is gained by 'undressing'. This is the case for high $T_c$ cuprates, where the 'dressing'
in the normal state is most apparent. However the same physics applies to all superconductors, whether
high, medium or low $T_c$\cite{bondc}. The superfluid electrons are undressed, paired, and otherwise completely free, so that
 they behave very much like electrons in diamagnetic atoms. The difference between microscopic atoms and
'giant atom' superconductors is the same as that between Rutherford atoms and 'Thomson atoms'\cite{thomson}.

In our previous work we have stressed the dressing   of carriers due to the
electron-electron interaction. However, by changing the mass of the electron from its bare free electron value, the
electron-ion interaction $also$ dresses the bare electron, increasingly so as the Fermi level moves up in the band. 
Realizing that 'undressing' involves undressing from $both$
the electron-electron $and$ the electron-ion interaction leads to the giant atom scenario\cite{atom}.

\section{Dressing from the electron-ion interaction}
When an electronic energy band is filled from bottom to top, the electrons that come in first are free to choose the state
that best suits them. Also, when only few electrons exist in the band they will adjust their state to take
maximum advantage of the crystal potential, without being much affected by interaction with each other.
These happy 'bonding electrons' have a low energy, a smooth wave function with large amplitude between ions
to take maximum advantage of the electron-ion potential while minimizing their kinetic energy, and an effective mass 
not very different from the
free electron mass. Each of them contributes  to the electrical conductivity of the metal, 
to the low frequency optical conductivity, and to the cohesion of the solid by giving rise to an effective 
attraction between ions (hence the name 'bonding').

\begin{figure}
\includegraphics[height=.25\textheight]{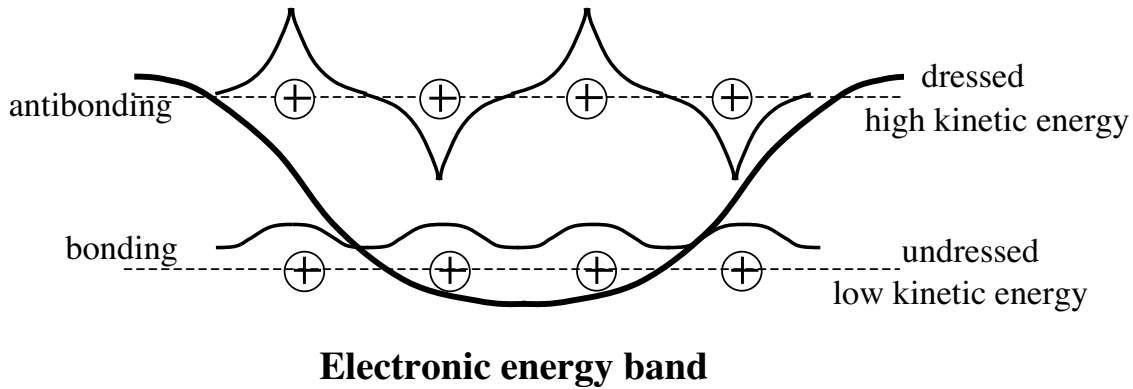}

  \caption{Electronic states in a band. The states at the bottom (bonding) have a high density of charge 
in-between the ions, and a  smooth wave function. The states at the top (antibonding)
 have a node in the charge density between the ions and a spiky wave function.}
\end{figure}

The situation is very different for the unhappy electrons that come in near the end of the
band-filling process, the 'antibonding electrons'. Because of the Pauli principle these electrons cannot take full
advantage of the electron-ion potential, since they have to be orthogonal to the pre-existing bonding
electrons. Hence their wavefunction has to oscillate greatly, having maximum amplitude at the ionic sites
and vanishing in-beween,
as shown schematically in Figure 1. With such constraints in fact they would rather reside on
isolated ions, which is why they give rise to a repulsive force between the ions (hence the name
'antibonding'), and sometimes succeed in breaking up the solid or at least driving it into another
more stable configuration (which is why lattice instabilities are associated with the existence of these
electrons at the Fermi energy). The large oscillations in the wave function give rise to a large kinetic energy.

Furthermore the antibonding electrons do not contribute to the electrical conductivity of the solid nor to the
low frequency optical conductivity, as a matter of fact they 'anticontribute'. When an external force
is applied to the electron (eg by applying an electric potential difference to the solid) both the electron
and the ionic lattice will pick up the added mechanical momentum:
\beq
F\Delta t=\Delta p_{el}+\Delta p_{latt}
\eeq
According to semiclassical transport theory,
\beq
F=\hbar \dot{k}
\eeq
\beq
\Delta p_{el}=m_e\Delta v_{el}=\frac{m_e}{\hbar}\Delta (\frac{\partial \epsilon_k}{\partial k})
=m_e\ (\frac{1}{\hbar^2}\frac{\partial^2\epsilon_k}{\partial^2k})\ \hbar\dot{k}\Delta t=\frac{m_e}{m^*}F\Delta t
\eeq
with $m^*$ the effective mass, which is $negative$ for antibonding electrons,
and $m_e$ the free electron mass. Hence the electron acquires
a momentum that is $opposite$ to the applied force, and the ionic lattice picks up the difference:
\beq
\Delta p_{latt}=(1-\frac{m_e}{m^*})F\Delta t
\eeq

If we quantify the 'dressing' of the free electron by the momentum transfer to the lattice Eq. (4), it is clear that
the electron-ion interaction increasingly dresses the electron as the Fermi level goes up in the band.
This is clearly seen in 'nearly free electron' theory, i.e. perturbation theory, where the band energy is given by
\beq
\epsilon_k=\epsilon_k^0+\sum_K\frac{|U_K|^2}{\epsilon_k^0-\epsilon_{k+K}^0}
\eeq
with $\epsilon_k^0=\hbar^2k^2/2m_e$ the free electron energy and $U_K$ the electron-ion
Fourier component for reciprocal lattice vector $K$. The effective mass is close to the free electron mass
near the bottom of the band ($k=0$) where the energy denominator in Eq. (5) is largest. As $k$ increases
the effective mass increases, then diverges and becomes negative and decreases in magnitude. The
momentum transfer to the lattice Eq. (4) increases monotonically throughout this process. Beyond
weak coupling perturbation theory, it is clear that this will be qualitatively similar for all energy bands
since $m^*$ is positive near the bottom and negative near the top.

Similarly the electrical conductivity $per$ $electron$ as well as the Drude weight per electron decrease
monotonically as the Fermi level rises. The Drude weight is given by
\beq
\frac{2}{\pi e^2}\int_{intraband} d\omega \sigma_1(\omega)=\frac{n_e}{m^*}
\eeq
The equality holds for the Fermi level close to the bottom of the band, with $n_e$ the number of electrons. Near the top
of the band, $n_e$ is replaced by the number of holes , $n_h=2-n_e$, and $m^*$ in Eq. (6)
by its absolute value. The difference between the Drude weight Eq. (6) and 
$n_e/m_e$, the corresponding value for free electrons, also quantifies the amount of 'dressing' and 
this difference increases
as the Fermi level rises from the bottom to the top of the band. It also represents
the optical spectral weight that is transfered from low intra-band frequencies to higher inter-band 
frequencies due to the electron-ion interaction.

Superconductivity can be understood as originating  in the desire of antibonding electrons at the Fermi
energy to contribute rather than 'anticontribute' to the electrical conductivity. To do so they have
to become like the bonding electrons at the bottom of the band, 
getting around the fact that  those single particle states
are already occupied. By pairing the antibonding electrons will be able to avoid the Pauli
principle, however to do so, it is necessary to recourse to the electron-electron interaction. This will be easiest
for the antibonding electrons when the influence of the electron-ion interaction force is small, which
is the case when the ions are negatively charged\cite{hole93}.

\section{ Dressing from the electron-electron interaction}

Another  essential aspect of 'dressing' arises from the effect of the electron-electron interaction.
We have dealt with this aspect extensively in our published work on hole superconductivity
and hence will only discuss it briefly  here.
Dressing from the electron-electron interaction parallels  the 
dressing from the
electron-ion interaction discussed above. To separate the two aspects, consider a tight binding band structure 
in a hypercubic lattice with nearest neighbor hopping only, so that the band effective mass has the same magnitude
for carriers at the bottom and the top of the band. We ignore here the aspect of dressing associated with the
$sign$ of the effective mass discussed above. When including the electron-electron interaction we also find that 
the dressing increases as the Fermi level goes up in the band, and results in a situation where holes
(carriers at the top of the band)  are highly dressed and electrons (carriers at the bottom of the band) are undressed\cite{elechub}.
Here the 'dressing' manifests itself in both an increase in the effective mass $m^*$ and a decrease in the
quasiparticle weight $z$, which are approximately related by $m^*=1/z$. 
The many-body Hamiltonian that describes this physics is a dynamic Hubbard model\cite{dynhub}, and its projection to
a low energy effective Hamiltonian yields a Hubbard model with correlated hopping\cite{correlh}
, with the hopping amplitude
decreasing as the Fermi level rises in the band.

\begin{figure}
  \includegraphics[height=.25\textheight]{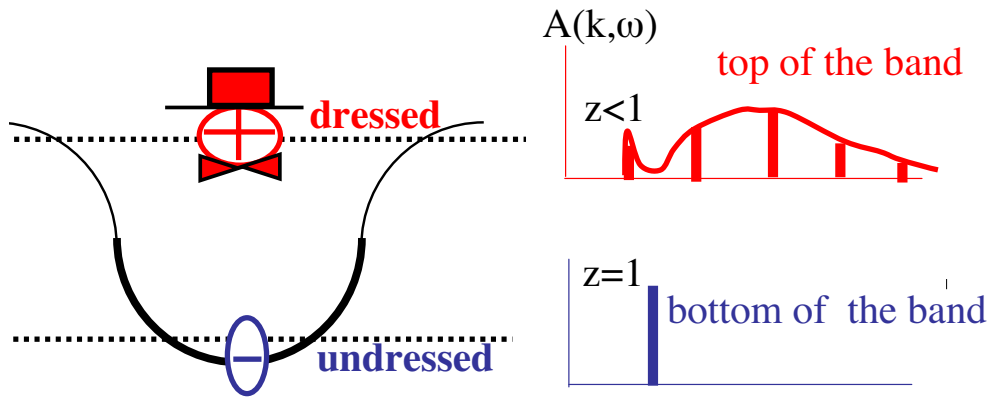}
  \caption{Electronic states in a band. The states at the bottom have a spectral function $A(k,\omega)$
with a single quasiparticle
peak of weight $z=1$.The states at the top of the band have a spectral function with a small quasiparticle peak
($z<1$) and a broad incoherent spectrum at higher energies.}
\end{figure}

\section{Superconductivity from undressing: experimental evidence}
We have argued in the previous sections that as the Fermi level goes up in the band the carriers become
increasingly dressed by $both$ the electron-ion and the electron-electron interaction.
Experiments show that in superconductors carriers are dressed  in the normal state, that 
the superconducting transition involves 'undressing', and that carriers are completely
undressed  in the  superconducting state, as follows:
\subsubsection{ Evidence that carriers in the normal state are dressed }
A look at the periodic table shows that essentially all superconducting elements have positive Hall coefficient,
indicating hole transport, or equivalently large 'dressing' of the bare electron by the
electron-ion interaction. The same is true for
compounds\cite{chapnik}.
  In the highest $T_c$ materials, the cuprates, carriers in the normal state are so heavily dressed
that the validity of the quasiparticle concept has been called into doubt, especially in the underdoped regime. This is seen from photoemission and optical conductivity measurements, that show small amplitude for
low frequency coherent response and large amplitude for high frequency incoherent response in
the normal state, indicating small quasiparticle weight and large effective mass of carriers in the
normal state, i.e. large 'dressing'\cite{ding,uchida}.
Upon hole doping the
normal state carriers become less dressed, consistent with the fact that the Fermi level goes down in the
band and  dressing due to both electron-electron and electron-ion interaction decreases as discussed
above. The fact that superconductors are often close to lattice instabilities indicates the presence
of 'antibonding' electrons, i.e. electrons 'dressed' by the electron-ion interaction, and the fact that
superconductors are usually poor conductors of electricity in the normal state indicates
'dressing' of normal state carriers.

\subsubsection{ Evidence that the superconducting transition involves undressing}
This is most clearly seen in systems where dressing is highest in the normal state. Photoemission experiments in
cuprates show the emergence of a quasiparticle peak as the system goes superconducting 
(increase in $z$)\cite{ding}.
The effect is greatest in the underdoped regime where the dressing in the normal state is
highest, as expected. This indicates that dressing is suppressed
by the transition to superconductivity and indirectly that the effective mass decreases through the
relation $m^*=1/z$. Optical experiments show directly that the effective mass decreases, as they show a
transfer of spectral weight from high to low frequencies as a system 
goes superconducting\cite{marel,santander,blumberg}.
Concerning undressing from the electron-ion interaction, we argue that this is evidenced by
the observed sign reversal of the Hall coefficient below $T_c$, that  turns from positive to negative before it 
goes to zero\cite{reversal}. This suggests that  the dressing from the electron-ion interaction that caused the
Hall coefficient to be positive in the normal state is eliminated when going superconducting.

\subsubsection{ Evidence that carriers in the superconducting state are undressed }
The 'Bernoulli potential',  the electric potential that develops between regions in the
superconductor where the superfluid is moving with different velocities reveals the sign of the
charge carriers. All experiments performed indicate that its
sign is consistent with negative electrons being the charge carriers\cite{bernoulli1}
, even though conventional theory
predicts its sign should be the one corresponding to the sign 
of the charge carriers in the normal state\cite{bernoulli2}.
The measured magnitude is also consistent with carriers in the superconducting state being free
electrons. The  sign and magnitude of the magnetic field that exists in the interior of a rotating superconductor (London field)\cite{londonfield} indicate that the carriers of the
superfluid current are undressed electrons\cite{londonfieldmy}. This is also indicated by the sign and magnitude of
the gyromagnetic effect, the change in angular momentum of a superconducting body
that occurs when an external magnetic field is applied\cite{gyro}. 

 \section{ How is undressing related to pairing and  superconductivity?}
 We know that dressing increases when the Fermi level rises in the band, and is largest with the band
almost full when the carriers are holes. When holes pair, the band becomes locally less full, hence undressing should
occur since locally  the Fermi level has moved down to where the carriers are less dressed. 
And of course we know
from BCS theory that pairing is associated with superconductivity. The dynamic Hubbard model 
predicts that pairing and superconductivity should occur at low $T$ because undressing causes a lowering of
kinetic energy, hence a decrease of free energy at low $T$ compared to the normal state\cite{dynhub}
. The correlated hopping
term in the low energy effective Hamiltonian shows clearly how the effective mass and the kinetic energy
decrease when the system goes superconducting\cite{correlh}. The dynamic Hubbard model also shows that the
quasiparticle weight increases and that transfer of spectral weight from high to low frequency occurs
upon pairing which is a signature of undressing\cite{qpundr}.

\section{ Historical precedents}
Our theory involves pairing as BCS theory does, and in addition it incorporates many ideas that
were discussed earlier but were   completely abandoned after BCS theory became established.

The idea that superconductors exhibit quantum mechanics at a macroscopic scale became generally accepted
around the mid 1930's after the Meissner effect was discovered and London's phenomenological
theory proposed. With it came the idea that electronic wavefunctions in superconductors extend 
coherently over the entire
macroscopic body. This naturally follows from considering the expression for the atomic diamagnetic susceptibility
\beq
\chi_0=-\frac{e^2}{6m_ec^2}<r^2>
\eeq
which is small when $<r^2>$ is of atomic dimensions but grows without bounds for macroscopic samples.
When the sample becomes big the effect of the magnetic field generated by the electrons themselves
becomes important and the susceptibility is
\beq
\chi=\frac{\chi_0}{1-4\pi \chi_0} \rightarrow  - \frac{1}{4\pi}
\eeq

The idea that the electronic wavefunction extends coherently over the entire superconducting
body is an essential ingredient of the 'giant atom' concept. In fact, after Meissner's discovery and London's theory it was quite common to refer to superconductors as being analogous to 'big atoms' 
or 'giant atoms'\cite{london}. This was by reference to Eq. (7) and
diamagnetism. However other possible consequences of the 'giant atom' concept like inhomogeneous
distribution of positive and negative charge, as in atoms, or the possibility of spin-orbit coupling, were not considered at that time.

 The idea of
a 'non-viscous electronic fluid' describing the superconducting electrons 
was popular at that time\cite{becker}, implying
no momentum transfer but rather a complete detachment between electrons and lattice. 
Specifically, Kronig\cite{kronig}
proposed to ignore the discrete ionic potential altogether and replace it by 'jellium'. This  picture  is inconsistent
with antibonding states at the top of the band which transfer large momentum to the ionic lattice, and suggests instead that the superfluid carriers behave more like electrons at the
bottom of the band. At the same time the fact that superconductors in the normal state exhibit hole rather than
electron transport was well known and discussed at the time\cite{holeold}, so the connection between transition
to superconductivity
and 'undressing' from the electron-ion interaction was somehow implicit in the discussions back then. Of course the concept of pairing was not being discussed at the time. Also the idea that superconductivity
may involve a reduction of the carrier's effective mass was discussed (by no other than Bardeen!)
in the pre-BCS era\cite{bardeen}.

As another precedent worth citing, Meissner wondered whether the electrons that carry the supercurrent
are the same electrons that carry the normal current, or whether new carriers become available
in the superconducting state\cite{meissner}. He favored the latter, based on the observation that superconducting
elements have more than one electron outside closed shells. This is in qualitative
agreement with the principle
discussed here: antibonding electrons 'anticontribute' to electrical conductivity in the normal
state, and they carry the supercurrent after they undress in the superconducting state.

\section{ The rotating superconductor puzzle}
A simply connected superconducting body rotating with angular velocity $\vec{\omega}$ has a magnetic field
\beq
\vec{B}=-\frac{2m_ec}{e}\vec{\omega}
\eeq
throughout its interior\cite{londonfield}, with $m_e$ the $free$ electron mass. 
Its origin when a superconducting body is put into rotation can be understood
as follows\cite{becker}: the rotating positive ions generate an electric field through Faraday's law:
\beq
\oint \vec{E}\cdot \vec{dl}=-\frac{1}{c} \frac{d}{dt}\int \vec{B}\cdot\vec{ds}
\eeq
hence for a constant magnetic field the electric field at distance r from the axis of rotation is
\beq
E=-\frac{r}{2c}\frac{d B}{dt}
\eeq
and the superfluid changes its velocity according to 
\beq
m_e\frac{dv_s}{d t}=eE=-\frac{er}{2c}         \frac{dB}{dt}
\eeq
For the superfluid initially at rest the final velocity is then
\beq
v_s=-\frac{er}{2m_ec}B
\eeq
In the interior the superfluid rotates together with the lattice, so that if the body is rotating
with angular velocity $\omega$, $v_s=\omega r$ and
\beq
\omega = -\frac{e}{2m_ec} B
\eeq
from which Eq. (9) follows. Alternatively, Eq. (9) can also be directly derived from the London equation.

There is a problem with this however. The centripetal force required for an electron to rotate
with angular velocity $\omega$ at radius $r$ is only one-half the one provided by the Lorenz
force due to the magnetic field Eq. (9). In other words, an electron in a magnetic field
$B$ rotates at the cyclotron frequency $\omega=eB/m_ec$ rather than at the Larmor
frequency Eq. (14). The giant atom scenario resolves this difficulty as explained below\cite{last}.

\section{The giant atom scenario}
If we accept that the wavefunction of the superconducting electron extends coherently over 
macroscopic distances  it is
difficult to imagine how it could 'know' about the microscopic variation of the ionic potential over the scale of Angstrom and
adjust accordingly, as antibonding electrons have to do. Rather, it is natural to conclude 
that the antibonding electrons at the Fermi energy 
manage to detach themselves completely from the lattice ('undress' from the electron-ion interaction)
and adopt a 'long wavelength' wavefunction that does not 'see' 
microscopic details. In that case each of these electrons
 will see a smeared positive charge distribution, screened self-consistently by
all the other electrons in the system. We call this the 'giant atom' scenario\cite{atom} (Figure 3).

\begin{figure}
  \includegraphics[height=.3\textheight]{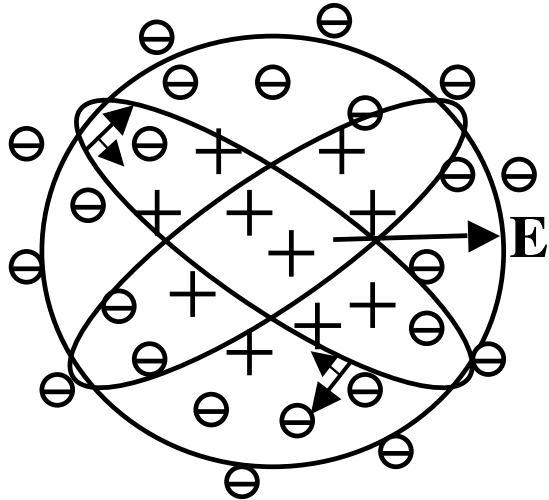}
  \caption{Superconductor as a giant atom (schematic). Electrons are pushed out of the interior towards the surface.
An electric field pointing out exists in the interior of the superconductor.
Some electrons will 'spill out' outside the surface of the superconducting body\cite{atom}.}
\end{figure}

In microscopic atoms, electrons do interact with each other, yet the response to a magnetic field 
involves the free electron mass. Similarly we argue that superfluid
electrons in the giant atom superconductor respond with
their free mass, as evidenced by Eq. (9), even though they interact with other electrons. This follows from galilean invariance.  It is only if it can see the discrete non-translationally invariant
ionic potential that the electron can respond with an effective mass different from the bare mass,
since only in that case can the lattice pick up some momentum.  

For electrons in filled shells of atoms, the change in velocity upon application of a magnetic field
is\cite{slater}
\beq
\Delta v=-\frac{e}{m_ec} A
\eeq
with $A$ the magnetic vector potential. This
follows from microscopic quantum mechanics, as well as from simple classical arguments: 
in an atom, 
the centripetal acceleration
for the electron's orbit is provided by the electric field from the ionic charge, $E_{ion}$
\beq
\frac{m_ev^2}{r}=eE_{ion}
\eeq
and the change in centripetal acceleration upon application of a magnetic field $B$ is provided
by the magnetic Lorenz force
\beq
\frac{2m_e v \Delta v}{r}=\frac{ev}{c}B
\eeq
from which Eq. (15) follows for $\vec{A}=\vec{B}\times\vec{r}/2$.
For a rotating superconductor, Eq. (9) for the magnetic field in its interior is consistent with the
relation Eq. (17) for $\Delta v=\omega r$. The derivation of Eq. (17) from Eq. (16) is only valid
if $\Delta v<<v$ and indicates how to resolve the puzzle raised by the magnetic field in the
rotating superconductor: {\it the electrons need to be rotating already (as in the ordinary atom) at a speed
much larger than $\omega r$ before the superconductor is set into rotation.}

However, this can only be possible if there is an electric field inside the superconductor  that provides their centripetal acceleration in the absence
of magnetic field, as in the case of the atom Eq. (16). This implies that the electronic charge distribution
in the superconductor,
just as in the atom, cannot be homogeneous, rather there has to be more
negative charge near the surface and more positive charge
in the interior. But that is precisely what the model of hole superconductivity predicts, that the
superconductor   expels negative charge from its interior\cite{chargeimb}.

\begin{figure}
  \includegraphics[ width=15cm]{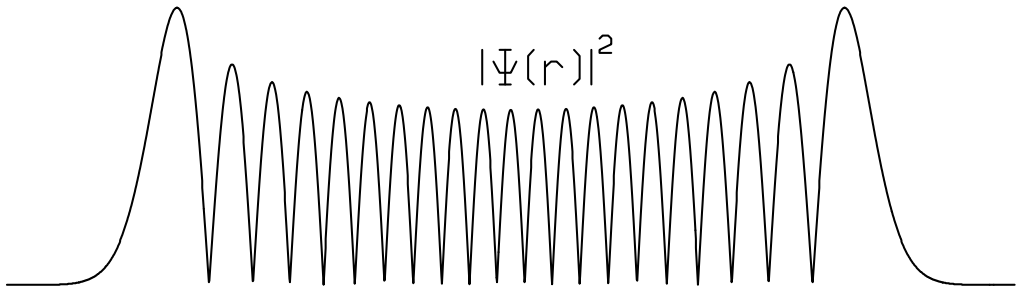}
  \caption{Wave function of harmonic oscillator for large quantum number. The wave function amplitude is largest
near the region of maximum classical elongation. As a consequence the electronic density is not uniform, in contrast
to the plane-wave free-electron model.}
\end{figure}

These considerations together with the 'undressing' phenomenology lead us to the model of a
 macroscopic 'Thomson atom', where a positive uniform charge distribution
exists over the volume of the superconductor\cite{atom}, except within a penetration length of 
the surface where the compensating negative charge resides. Consider 
a spherical or cylindrical geometry for simplicity. The electric
field from a uniform charge distribution is linear in r, and the potential acting on the electron is a harmonic oscillator
potential. A wavefunction for such a potential for large quantum number is shown in Figure 4  . 
It is not uniform but has
larger amplitude for large r. This is of course obvious from the fact that an oscillator spends most of its time near
the region of maximum elongation, and justifies self-consistently the assumption that the electron sees a positive charge
distribution in the interior.

Of course the giant atom scenario with an electric field in the interior is not enough. Electrons in such a potential, even in the absence of imperfections, will not exhibit the Meissner
effect but rather the much weaker Landau diamagnetism. This is because there will be a rearrangement
when a magnetic field is applied\cite{welker}: electrons that
are orbiting with their angular momentum parallel to the applied field will increase their velocity, and would lower their energy
by turning around and joining the electrons orbiting with angular momentum antiparallel to the applied 
magnetic field which are slowed
down by it. This is where the pairing condition comes in: if for every electron with $k\uparrow$ there is a partner
with $-k\downarrow$ and it costs a finite energy $\Delta$ to break up the pair, it will prevent (unless the magnetic
field is strong enough) the electron from 'turning over' 
because that would break the pair. So the BCS energy gap provides the necessary 'rigidity' for the pair to respond
like London rather than Landau electrons.

The pairing requires that for every $k\uparrow$ electron there is a partner $-k\downarrow$ electron, but it $does$ $not$
require that there be a corresponding $-k\uparrow$, $k\downarrow$ pair. In fact we argue that this will quite generally
not be the case, due to spin-orbit coupling. The energy of an  electron in a macroscopic orbit will be lower
if its orbital angular momentum is antiparallel to its spin, because in its restframe it sees a magnetic field from the
positive background that is parallel to its spin magnetic moment. Thus we argue that spin up electrons will be orbiting
predominantly in one direction, and spin down electrons in the opposite direction, 
as shown  schematically in Fig. 3, giving rise to a macroscopic
spin current which, as explained above, is fully compatible with the pairing concept.

From a purely classical point of view, such a spin current is inescapable if we accept the concept that an outward pointing
electric field exists in the interior of superconductors and that the negative charge density is higher near the surface. In order
for a superfluid  electron at radius r to be in mechanical equilibrium it needs to be rotating so that $m_ev^2/r$ equals the
force from the electric field pulling it in. In the absence of magnetic field and rotation of the superconductor there is no
charge current field $v_s(r)$ of the superfluid in the interior of the superconductor. But there can and will  be a spin current field
$v_{s\sigma}(r)$   such that $v_{s\sigma}(r)=-v_{s,-\sigma}(r)$.

Some experimental consequences of the giant atom scenario were discussed 
in Ref.\cite{atom}. We predict that the spin current
should give rise to a quadrupolar electric field around superconducting rings, that should be experimentally
detectable, and to forces between superconducting rings due to these electric fields. Furthermore we predict that just
as in regular atoms the negative charge will not be confined to the region of the positive charge but rather will
leak out of the superconducting body, of course in much smaller proportion than for a regular atom. 
This is likely to play a role in the proximity effect\cite{proxi}. Finally, since in the normal state the negative
charge distribution is homogeneous we predict that a radially outward electron current is generated when the system goes
into the superconducting state, to generate the nonhomogeneous charge distribution characteristic of the
superconductor. The interaction of this radial current with the ionic background will deflect up- and down-spin electrons
tangentially in opposite directions to give rise to a spin current of precisely the sign discussed above through the
mechanism discussed in Ref.\cite{hallferro}. Furthermore, the radially outgoing electron current in the presence
of a magnetic field will give rise to screening currents that will cancel the magnetic field in the
interior of the superconductor as required by the Meissner effect.

The concept of an electric field in the interior of the superconductor however raises a question: why isn't it
screened by mobile charges, as electric fields are screened in the interior of normal metals? For the superfluid 
electrons we have
argued that the electric field produces a force that is the centripetal force that sustains the  superfluid spin current. However at finite temperature there
will also be thermally excited quasiparticles which can move . Why is it that they do not
screen the electric field in the interior and nullify it?

This question  in fact also arises in the conventional framework when the superconductor is rotating, since
an electric field needs to exist to balance the forces on the rotating superfluid electrons, and
has never been addressed in that framework. For our case there appears to be a simple answer: the theory of hole superconductivity
predicts that quasiparticles are $positively$ $charged$\cite{thermo}, with charge given by
\beq
Q_k=e(u_k^2-v_k^2)
\eeq
which is positive on the average due to electron-hole asymmetry. Positive quasiparticles will be pushed $out$ by the
electric field inside superconductors and not screen the field. In equilibrium a density gradient of quasiparticles will be
established so that the electrochemical potential for quasiparticles will be constant throughout the volume of
the superconductor.

\section{Conclusion}
The theory considered here
 proposes that superconductors in the normal state have carriers at the Fermi energy
that are highly dressed by both electron-ion
and electron-electron interactions, and that when a metal becomes superconducting
 these carriers
completely undress from both electron-ion and electron-electron interactions and become  
free-electron-like, except for the pairing correlations. Superconductivity is driven by electron
undressing, and the resulting system is a macroscopic
'Thomson atom' with paired time-reversed electrons.
The facts that the theory may lead to an understanding of a variety of puzzling phenomena
in superconductors\cite{last}, that
it provides a unified scheme to explain superconductivity in all superconductors\cite{narlikar}, that it has
a plausible microscopic foundation\cite{holeelec}, and that it can be understood in terms of a single
underlying physical principle, undressing, argue for its validity. Ultimately
its validation or refutation will come from experiments testing its predictions. Various of its
experimental predictions are consistent with observations as discussed in the references.

\begin{acknowledgments}
The author is grateful to the organizers and participants of the meeting
"Highlights in Condensed Matter Physics", Salerno, Italy, May 9-12, 2003, for the opportunity to present
and discuss these concepts in a stimulating environment.
\end{acknowledgments}



 
\end{document}